\documentclass[aps,pra,twocolumn]{revtex4}
\usepackage{amsmath}
\usepackage{amssymb}
\usepackage{dcolumn}
\usepackage{graphicx}

\def\be{\begin{equation}}
\def\ee{\end{equation}}
\def\e#1{\label{#1}\end{equation}}
\def\bea{\begin{eqnarray}}
\def\eea{\end{eqnarray}}
\def\ea#1{\label{#1}\end{eqnarray}}

\def\bem#1{\begin{mathletters}\label{#1}}
\def\eml{\end{mathletters}}

\def\ket#1{{|#1\rangle}}

\def\4#1{{\boldsymbol{#1}}}
\def\8#1{{\widetilde{#1}}}

\begin{document}

\title{Unity fidelity multiple teleportation using partially entangled states}

\author{Gustavo Rigolin}
\email{rigolin@ufscar.br} \affiliation{Departamento de Fisica,
Universidade Federal de Sao Carlos, Caixa Postal 676, Sao Carlos,
13565-905, SP, Brazil}

\begin{abstract}
We show that the 
multiple teleportation protocol (MTP)
given in Ref. [Phys. Rev. Lett. \textbf{100}, 110503 (2008)] is
not restricted to the Knill-Laflamme-Milburn (KLM) framework.
Rather,
we show that MTP 
can be implemented using
any teleportation scheme. 
We also present two new
MTP's which, under certain situations, are more efficient than the original one, 
requiring 
half of the 
number of its teleportations 
to achieve at least the same probability of success
($\mathcal{P}_{suc}$). One of the protocols, however, uses 
less entanglement than 
the others yielding, surprisingly, the greatest
$\mathcal{P}_{suc}$.
\end{abstract}

\maketitle

\section{Introduction}

The importance of quantum teleportation \cite{Ben93} is widely
recognized today. Not only does it enable the remote transmission
of the state describing a quantum system to another one, without
ever knowing the state, but it also allows the construction of a
new way to perform quantum computation \cite{Got99,Kni01}. In the
previous and in many other applications of teleportation, it is
desirable, if not crucial, that the teleported state arrives at
its destination (Bob) exactly as it left the preparation station
(Alice). In other words, we want a unity fidelity output state,
which is always achieved
if Alice and Bob share a maximally entangled state (MES)
\cite{Ben93}.
However, there might happen that our parties do not share a MES
or, in addition, intermediate teleportations to other parties must
be done before the state reaches Bob. This limitation can be
overcome by distilling out of an ensemble of partially entangled
states (PES's) maximally entangled ones \cite{Ben96}. But this
approach requires a large amount of copies of PES's to succeed and
is ineffective when just a few copies are available. Another way
to achieve unity fidelity teleportation with limited resources is
based on the probabilistic quantum teleportation (PQT) protocols
of Refs. \cite{Agr02,Gor06,Guo00}.

Recently, in an interesting work, Mod{\l}awska and Grudka
\cite{Gru08} presented yet another way of achieving
probabilistically unity fidelity teleportation. Their strategy was
developed in the framework of the KLM scheme \cite{Kni01} for
linear optical teleportation. The main idea behind their approach
was the recognition that multiple (successive) teleportations
using the \textit{same} PES increased the chances of getting a
perfect teleported qubit. We can also see the ideas of Ref.
\cite{Gru08}, as generalized here, as a way to extend the
usefulness of quantum relays \cite{Bri98} whenever non MES's are
at stake and entanglement concentration is not practical (only a
few copies of entangled states are available).

In this contribution we show that the features of the MTP of Ref.
\cite{Gru08} are not restricted to the KLM teleportation scheme.
In order to show that we build in Sec. \ref{MTPs} a similar
protocol (protocol $1$) without relying on the intricacies of the
KLM scheme. Actually, we use the same language of the original
Bennett \textit{et al.} proposal \cite{Ben93}, which allows us to
express $\mathcal{P}_{suc}$, the total probability of getting
unity fidelity outcomes, as a function of the number of
teleportations and of the shared entanglement between Alice and
Bob.  We then present two new protocols (protocols $2$ and $3$,
see Fig. \ref{Fig1}), both of which are more efficient than the
previous one. An important feature of these protocols is that they
give $\mathcal{P}_{suc}>1/2$ for a huge class of PES's. This is
particularly useful when we have a few copies of the qubit to be
teleported, since after a few runs of the MTP the overall
$\mathcal{P}_{suc}\rightarrow 1$. On top of that, protocol $2$
possesses the same efficiency of the first one but needs only
\textit{half} the number of teleportations to achieve the same
$\mathcal{P}_{suc}$. We also show that this protocol is connected
to the PQT of Refs. \cite{Agr02,Gor06}. Protocol $3$, on the other
hand, in addition to requiring just half the number of
teleportations of protocol $1$ also achieves the highest
$\mathcal{P}_{suc}$. Actually, we show that for some set of PES's
$\mathcal{P}_{suc} \approx 1$ after just a few teleportations
within a single run of the MTP. Moreover, and surprisingly, at
each successive teleportation this last protocol requires less and
less entanglement to properly work. In Sec. \ref{comparison} we
compare the efficiencies of all the three protocols presented here
with a different strategy to achieve unity fiedelity teleportation
based on entanglement swapping \cite{Bos99}. In particular, we
compare our results with those obtained for multiple entanglement
swapping as presented in Ref. \cite{Per08}. We show that, under
certain conditions, we can achieve a better performance using the
protocols here presented.

\section{Multiple teleportation protocols}
\label{MTPs}

\begin{figure}[!ht]
\includegraphics[angle=0,width=7cm]{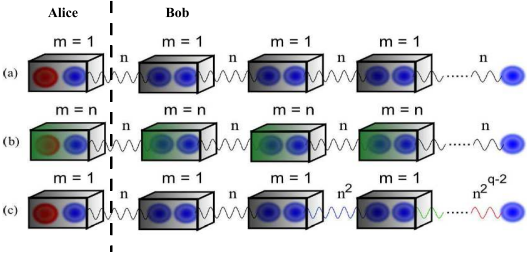}
\caption{\label{Fig1}(Color online) Pictorial view of all MTP's
after $q$ teleportations. Note that all but the first PES is
shared between Alice and Bob. All the others are at Bob's. (a)
Protocol $1$: Boxes denote standard BM's ($m=1$) and at each
teleportation the quantum channel is the same state
$\ket{\Phi^+_n}$. (b) Protocol $2$: Boxes now denote GBM's with
$m=n<1$ with the same state $\ket{\Phi^+_n}$ at each stage. (c)
Protocol $3$: Boxes denote standard BM's ($m=1$) but the quantum
channel's entanglement are successively reduced after the second
teleportation according to the following rule,
$\ket{\Phi^+_{n_j}}\rightarrow \ket{\Phi^+_{n^2_j}}$.}
\end{figure}

\textit{Protocol 1.} Let us assume that we have $j=N$ PES's
described by $|\Phi^+_{n_j}\rangle = f_{n_j}\ket{00} +
g_{n_j}\ket{11}$, with $f_{n_j}=1/\sqrt{1+n_j^2}$ and
$g_{n_j}=n_j/\sqrt{1+n_j^2})$. (See panel (a) of Fig. \ref{Fig1}.)
We assume the first PES is shared between Alice and Bob while the
remaining $N-1$ are with Bob. Without loss of generality we set
$0<n_j<1$ \cite{Gor06} and for this protocol also that $n_j=n$,
$j=1,\dots, N$ \cite{Gru08}, i.e, same entanglement at each
teleportation. We can also build a generalized Bell basis as
follows,
%
%
%
\begin{eqnarray*}
\ket{\Phi_{m_j}^{+}}  =  f_{m_j}\ket{00} + g_{m_j}\ket{11}, &
\ket{\Phi_{m_j}^{-}}  =  g_{m_j}\ket{00} - f_{m_j}\ket{11},\\
\ket{\Psi_{m_j}^{+}}  =  f_{m_j}\ket{01} + g_{m_j} \ket{10},&
\ket{\Psi_{m_j}^{-}}  =  g_{m_j}\ket{01} - f_{m_j}\ket{10},
\end{eqnarray*}
with $m_j=1$ being the original Bell basis and the choice for
protocol $1$. Alice wants to teleport the qubit
$\ket{\phi^A}=\alpha\ket{0}+\beta\ket{1}$  and at each step $j$ a
Bell measurement (BM) is implemented whose result is known to Bob
(See Fig. \ref{Fig1}). This information allows him to correct the
final state applying the proper unitary operations conditioned on
the results of each BM \cite{Ben93}, i.e, $I$ if the BM yields
$\ket{\Phi^+}$, $\sigma_z$ for $\ket{\Phi^-}$, $\sigma_x$ for
$\ket{\Psi^+}$, and $\sigma_z\sigma_x$ for $\ket{\Psi^-}$, where
$I$ is the identity and $\sigma_{z,x}$ the
standard Pauli matrices. 

\begin{figure}[!ht]
\includegraphics[width=7cm]{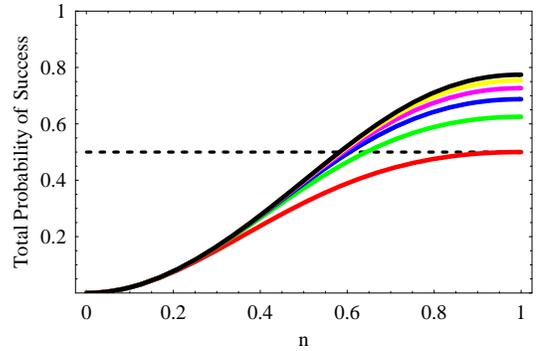}
\caption{\label{Fig2}(Color online) For protocol 1: From bottom to
top the curves represent $\mathcal{P}_{suc}$ after $q=2$, $4$,
$6$, $8$, $10$, and $12$ successive teleportations. For protocol
2: From bottom to top the curves show $\mathcal{P}_{suc}$ after
$q=1$, $2$, $3$, $4$, $5$, and $6$ successive teleportations. The
dashed curve shows the optimal probability (1/2) using the PQT
protocol. All quantities are dimensionless.}
\end{figure}

Before the first teleportation the state describing all qubits are
$\ket{\Phi}=\ket{\phi^A}\otimes_{j=1}^{N}\ket{\Phi^+_{n_j}}$,
which can be written as $\ket{\Phi}$ $=$ [$\ket{\Phi^+}$
$(f_1f_n\alpha\ket{0}$ $+$ $g_1g_n\beta\ket{1})$ $+$
$\ket{\Phi^-}$ $\sigma_z$ $(g_1f_n\alpha\ket{0}$ $+$
$f_1g_n\beta\ket{1})$ $+$ $\ket{\Psi^+}$ $\sigma_x$
$(f_1g_n\alpha\ket{0}$ $+$ $g_1f_n \beta\ket{1})$ $+$
$\ket{\Psi^-}$ $\sigma_z\sigma_x$ $(g_1g_n\alpha\ket{0}$ $+$
$f_1f_n \beta\ket{1})]$ $\otimes_{j=2}^{N}\ket{\Phi^+_{n_j}}.$
Unity fidelity teleportation occurs only if $f_1f_n=g_1g_n$ or
$f_1g_n=g_1f_n$. But this is only possible if we have a MES ($n=1$
$\rightarrow$ $f_n=f_1=g_n=g_1=1/\sqrt{2}$). Hence, after the
first teleportation $P^{(1)}_{suc}=0$. It is important to note
that at each teleportation, the previous teleported qubit is
changed to $\alpha_{j-1} \rightarrow \alpha_j = h_j^\alpha
\alpha_{j-1}$ and $\beta_{j-1} \rightarrow \beta_j = h_j^\beta
\beta_{j-1}$, with $(h_j^\alpha, h_j^\beta)$ $=$ $(f_{1}f_{n},
g_{1}g_{n})$, or $(g_{1}f_{n}, f_{1}g_{n})$, or $(f_{1}g_{n},
g_{1}f_{n})$, or $(g_{1}g_{n}, f_{1}f_{n})$, for $j>1$ and
$\alpha_0=\alpha$ and $\beta_0 = \beta$. We are neglecting
normalization for the moment. After the second teleportation there
exist $16$ possible outcomes ($4 \times 4$ pairs of BM's) for the
teleported qubit, which is described by one of $16$ states whose
coefficients are given by terms like $(\alpha_2,\beta_2)$ $=$
$(h_2^\alpha\alpha_1,h_2^\beta\beta_1)$ $=$ $(f_1f_n
f_1f_n\alpha,g_1g_n g_1g_n\beta)$, $(g_1f_n f_1f_n\alpha,f_1g_n
g_1g_n\beta)$, $\dots$, $(g_1g_n g_1g_n\alpha,f_1
f_nf_1f_n\beta)$. Of all possibilities, those giving unity
fidelity are such that $h_2^\alpha=h_2^\beta$, since we can factor
out the terms multiplying $\alpha$ and $\beta$ obtaining the exact
original state $|\phi^A\rangle$. To determine those successful
cases we first note that whenever $\ket{\Phi^{\pm}}$ is a result
of a BM the teleported coefficients change to $\alpha_{j}
\rightarrow  \alpha_{j}$ with $\beta_{j}\rightarrow n \beta_{j}$.
Second, whenever the BM results in $\ket{\Psi^{\pm}}$ we get
$\alpha_{j} \rightarrow n \alpha_{j}$ and $\beta_{j}\rightarrow
\beta_{j}$. Therefore, it is not difficult to see that we always
get unity fidelity teleportation when we have an equal number of
$|\Phi^{\pm}\rangle$ and $\ket{\Psi^{\pm}}$ in a sequence of BM's,
or equivalently, an equal number of functions $g_n$ multiplying
$\alpha$ and $\beta$. For the case of two teleportations the
successful cases are given by eight possibilities:
$\ket{\Phi^{\pm}}\ket{\Psi^{\pm}}$ and
$\ket{\Psi^{\pm}}\ket{\Phi^{\pm}}$. The probability of all those
cases are equal and is given by $P_{event}^{(2)}$ $=$
$n^2/[4(1+n^2)^2]$. Thus, $P_{suc}^{(2)}=2n^2/(1+n^2)^2$. If we
are successful, we do not need another teleportation. However, if
we fail, we need to proceed with successive teleportations, hoping
to get a balanced sequence of $|\Phi^{\pm}\rangle$ and
$\ket{\Psi^{\pm}}$ BM's. We can show that at the $q$-th
teleportation
\begin{equation}
P_{suc}^{(q)} = A(q)n^{q}/[2^q(1+n^2)^{q}], \label{A}
\end{equation}
where for $q$ odd $A(q)=0$ and for $q$ even A(q) is the number of
all possible combinations of $q$ BM's in which we have an equal
number of $|\Phi^{\pm}\rangle$ and $|\Psi^{\pm}\rangle$,
excluding, of course, those cases where we already had a balanced
number in the previous even teleportations. For the first $12$
teleportations we have $A(2)=8$, $A(4)=32$, $A(6)=256$,
$A(8)=2560$, $A(10)=28672$, and $A(12)=344064$. In Fig.~\ref{Fig2}
we plot the total probability of success after the $q$-th
teleportation, $\mathcal{P}_{suc}=\sum_{j=1}^q P_{suc}^{(j)},$ as
a function of $n$ (the greater $n$ the greater the entanglement).
Note that here and in the remaining of this section
$\mathcal{P}_{suc}$ is given by the sum of the probabilities of
all previous successful teleportations since the $N-1$ PES's are
with Bob. In Sec. \ref{comparison} we also study other scenarios,
in particular the one in which Bob possesses just one PES.
Looking at Fig. \ref{Fig2} we see that after each teleportation
$\mathcal{P}_{suc}$ increases at a lower rate. Also, after the
$10$-th teleportation we are already close to the maximal value of
$\mathcal{P}_{suc}$ for whatever value of $n$.  We should remark
that we are considering success only unity fidelity
teleportations. That is why $\mathcal{P}_{suc}$ does not tend to
one as $n\rightarrow 1$. Indeed, no matter how close $n$ is to one
we are always discarding the sequences of BM's where we do not get
a balanced set of measurements involving the Bell states
$|\Phi^{\pm}\rangle$ and $|\Psi^{\pm}\rangle$.
%

\textit{Protocol 2.} As before, we assume that one has $j=N$ PES's
described by $|\Phi^+_{n_j}\rangle$, with $0<n_j<1$ and $n_j=n$,
$j=1,\dots, N$. (See panel (b) of Fig. \ref{Fig1}.) However,
differently from protocol $1$, we now assume $m_j=m=n$, any $j$.
The state to be teleported is
$\ket{\phi^A}=\alpha\ket{0}+\beta\ket{1}$ and at each step $j$ one
implements a generalized Bell measurement (GBM)
\cite{Agr02,Gor06}. A GBM is a projective measurement of two
qubits onto one of the four generalized Bell states given above
(See Ref. \cite{Kim04} for ways of implementing a GBM.) The result
of each GBM is known to Bob who uses this information to apply the
right unitary operations on his qubit as described in the first
protocol. The rest of the present protocol is nearly the same as
before and is inspired by the PQT of Refs. \cite{Agr02,Gor06}.

Before any teleportation the state describing all qubits can be
written as $\ket{\Phi}$ $=$ [$\ket{\Phi^+_{m}}$
$(f_mf_n\alpha\ket{0}$ $+$ $g_mg_n\beta\ket{1})$ $+$
$\ket{\Phi^-_m}$ $\sigma_z$ $(g_mf_n\alpha\ket{0}$ $+$
$f_mg_n\beta\ket{1})$ $+$ $\ket{\Psi^+_m}$ $\sigma_x$
$(f_mg_n\alpha\ket{0}$ $+$ $g_mf_n \beta\ket{1})$ $+$
$\ket{\Psi^-_m}$ $\sigma_z\sigma_x$ $(g_mg_n\alpha\ket{0}$ $+$
$f_mf_n \beta\ket{1})]$ $\otimes_{j=2}^{N}\ket{\Phi^+_{n_j}}.$
Note that now we have rewritten the first two qubits using the
generalized Bell basis with $m=n$, i.e., we have imposed the
`matching condition', where the entanglement of the channel and of
the measuring basis are the same \cite{Agr02,Gor06}. This allows
us to obtain unity fidelity teleportation right after the first
teleportation whenever we measure $|\Phi^-_m\rangle$ or
$|\Psi^+_m\rangle$ with $P^{(1)}_{suc}=2n^2/(1 + n^2)^2$. The
previous step is precisely the PQT \cite{Agr02,Gor06}.
To analyze the other teleportations we need to keep in mind three
facts. (1) The $j$-th teleported qubit is changed to
$(\alpha_{j},\beta_j) \rightarrow (\alpha_{j}, n^2\beta_{j})$
whenever $\ket{\Phi^+_m}$ is a result of a GBM; (2) If the GBM
yields $\ket{\Phi^-_m}$ or $\ket{\Psi^+_m}$  we get
$(\alpha_{j},\beta_j) \rightarrow n(\alpha_{j},\beta_{j})$; (3) if
we measure $\ket{\Psi^-_m}$ the qubit goes to
$(\alpha_{j},\beta_j) \rightarrow (n^2\alpha_{j},\beta_{j})$.
Therefore, when we have an equal number of $|\Phi^{+}_m\rangle$
and $\ket{\Psi^{-}_m}$, $m=n$, in a sequence of GBM's we get unity
fidelity. The $n^2\beta_j$ coming from the measurement of
$|\Phi^{+}_m\rangle$ is compensated by the $n^2\alpha_j$ coming
from another GBM giving $\ket{\Psi^-_m}$. Note that the states
$\ket{\Phi^-_m}$ and $\ket{\Psi^+_m}$ are `neutral', giving an
overall $n$ that can be ignored for the determination of the
successful cases.
For example, after the second teleportation we have two possible
GBM outcomes where we have a unity fidelity teleportation, namely,
$\ket{\Phi^+_m}\ket{\Psi^-_m}$ and $\ket{\Psi^-_m}\ket{\Phi^+_m}$
with $P^{(2)}_{suc}=2n^4/(1 + n^2)^4$. And after the third
teleportation the successful cases are four:
$\ket{\Phi^+_m}\ket{\Phi^-_m}\ket{\Psi^-_m}$,
$\ket{\Phi^+_m}\ket{\Psi^+_m}\ket{\Psi^-_m}$,
$\ket{\Psi^-_m}\ket{\Phi^-_m}\ket{\Phi^+_m}$, and
$\ket{\Psi^-_m}\ket{\Psi^+_m}\ket{\Phi^+_m}$, with
$P^{(3)}_{suc}=4n^6/(1 + n^2)^6$.
In general, after the $q$-th teleportation we have,
\begin{equation}
P_{suc}^{(q)} = B(q)n^{2q}/[(1+n^2)^{2q}], \label{B}
\end{equation}
where B(q) is the number of all possible combinations of $q$ GBM's
where we have an equal number of $|\Phi^{+}_m\rangle$ and
$|\Psi^{-}_m\rangle$, excluding, as we did in protocol $1$, the
cases where we already got an equal number of those two states in
the previous teleportations. For the first six teleportations we
have $B(1)=2$, $B(2)=2$, $B(3)=4$, $B(4)=10$, $B(5)=28$, and
$B(6)=84$.

Noting that $A(2q)/2^{2q}=B(q)$ we immediately see that
Eqs.~(\ref{A}) and (\ref{B}) are the same. However, in protocol
$2$, we just need \textit{half} of the number of teleportations to
achieve the same efficiency, which is a quite remarkable economy
on entanglement resources. Also, the need for less teleportations
reduces other possible errors introduced by imperfect projective
measurements. Furthermore, this result connects the PQT of Refs.
\cite{Agr02,Gor06} to protocol $1$. This is true because two
successive teleportations using that protocol is equivalent to one
using protocol $2$, being the latter an extension of the PQT.

\textit{Protocol 3.} Like protocol $1$, here we do not need GBM's.
(See panel (c) of Fig. \ref{Fig1}.) The projective measurements
are made using the standard Bell basis, i.e., $m_j=1$, any $j$.
However, and differently from the previous protocols, we assume
that at each teleportation the entanglement of the quantum channel
is reduced according to the following rule: $n_{j}=n_{j-1}^2$,
$j\geq 3$ with $n_1=n_2=n<1$. In words, the first two
teleportations are done spending two entangled states
$|\Phi^+_n\rangle$ and after that we start using less and less
entanglement. The first two steps of this protocol are identical
to the first two of protocol $1$ yielding $P_{suc}^{(1)}=0$ and
$P_{suc}^{(2)}=2n^2/(1+n^2)^2$. After the second teleportation,
the unsuccessful cases are described by the state
$\alpha\ket{0}+n^2\beta\ket{1}$, if the BM's resulted in
$|\Phi^{\pm}\rangle|\Phi^{\pm}\rangle$, or by the state
$n^2\alpha\ket{0}+\beta\ket{1}$, if the two successive BM's
yielded $|\Psi^{\pm}\rangle|\Psi^{\pm}\rangle$. Since in the third
teleportation the entangled state spent is $\ket{\Phi^+_{n^2}}$,
the previous teleported qubit changes to $(\alpha_2,\beta_2)
\rightarrow (\alpha_{2}, n^2\beta_{2})$ if we measure
$|\Phi^{\pm}\rangle$ or to $(\alpha_{2},\beta_2) \rightarrow
(n^2\alpha_{2}, \beta_{2})$ if we get $|\Psi^{\pm}\rangle$. Hence,
whenever we get the following sequences of BM's,
$|\Phi^{\pm}\rangle|\Phi^{\pm}\rangle|\Psi^{\pm}\rangle$ or
$|\Psi^{\pm}\rangle|\Psi^{\pm}\rangle|\Phi^{\pm}\rangle$ we
achieve unity fidelity with
$P_{suc}^{(3)}=2n^4/[(1+n^2)^2(1+n^4)]$. The unsuccessful cases
are given by the following $16$ cases,
$|\Phi^{\pm}\rangle|\Phi^{\pm}\rangle|\Phi^{\pm}\rangle$ and
$|\Psi^{\pm}\rangle|\Psi^{\pm}\rangle|\Psi^{\pm}\rangle$, with the
unsuccessful teleported qubits being either $(\alpha, n^4\beta)$
or $(n^4\alpha, \beta)$, respectively.

It is now clear why we will use $\ket{\Phi^+_{n^4}}$ to implement
the fourth teleportation. We are trying to catch up with the $n^4$
that multiplies either $\alpha$ or $\beta$. And since the
unsuccessful cases after this step will turn to have a $n^8$
multiplying either $\alpha$ or $\beta$, we will need
$\ket{\Phi^+_{n^8}}$ at the fifth teleportation to catch up with
it. In general, after the $(q-1)$-th teleportation the
unsuccessful cases are those where we got the following sequences
of $q-1$ BM's: $\otimes_{j=1}^{q-1}\ket{\Phi^{\pm}}$ or
$\otimes_{j=1}^{q-1}\ket{\Psi^{\pm}}$, giving a total of $2\times
2^{q-1}$ cases with unsuccessful (not normalized) teleported
qubits described by $\alpha \ket{0} + n^{2^{q-2}}\beta\ket{1}$ or
$n^{2^{q-2}} \alpha \ket{0} + \beta\ket{1}$, respectively. At the
$q$-th teleportation we succeed if we have either
$(\otimes_{j=1}^{q-1}\ket{\Phi^{\pm}})\ket{\Psi^{\pm}}$ or
$(\otimes_{j=1}^{q-1}\ket{\Psi^{\pm}})\ket{\Phi^{\pm}}$ as our
sequence of BM's, with the probability to get any single
successful sequence being identical and given by
$P_{event}^{(q)}=n^{2^{q-1}}/[2^q(1+n^2)\prod_{j=1}^{q-1}(1+n^{2^j})]$.
But since we have a total of $2 \times 2^q$ successful sequences
we get ($q\geq 2$),
%
%
\begin{equation}
P_{suc}^{(q)}=2n^{2^{q-1}}(1-n^2)/[(1+n^2)(1-n^{2^{\,\!^q}})],
\end{equation}
where we used that $\prod_{j=1}^{q-1}(1+n^{2^{j}})= (1 -
n^{2^{q}})/(1- n^2)$. There is also a peculiar way of writing
$P_{suc}^{(q)}$ in terms of the concurrence \cite{Woo98},
$C_{n_j}=2n_j/(1+n_j^2)$, an entanglement monotone/quantifier for
the state $|\Phi^{+}_{n_j}\rangle$,
%
%
$$
P_{suc}^{(q)}=2\mbox{$\prod_{j=1}^{q}$}C_{n_j}/2, \hspace{1cm}
q\geq 2.
$$
Actually, for the other two protocols we can  write similar
expressions for $P_{suc}^{(q)}$. The difference comes from the
factor multiplying the product of concurrences. Here, this factor
is $2$, for the other protocols, they are $A(q)/2^{q}$ and $B(q)$.
In protocol $2$ we must also consider the concurrences of the GBM.
This changes, in the above expression for $P^{(q)}_{suc}$, the
term $C_{n_j}/2$ to $C_{n_j}C_{m_j}/4$.

\begin{figure}[!ht]
\includegraphics[width=7cm]{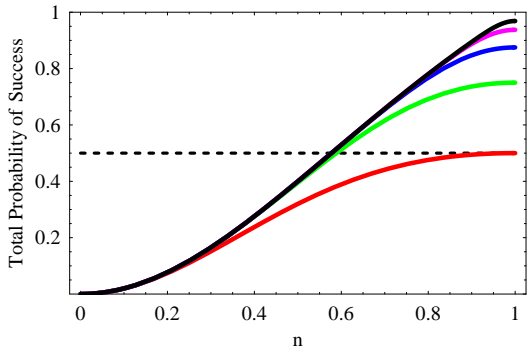}
\caption{\label{Fig3}(Color online) From bottom to top the curves
show $\mathcal{P}_{suc}$ after $q=2$, $3$, $4$, $5$, and $6$
successive teleportations using protocol $3$. The dashed curve
shows the optimal probability (1/2) obtained using the PQT
protocol.}
\end{figure}

In Fig. \ref{Fig3} we plot the total probability of success after
$q$ teleportations, $\mathcal{P}_{suc}=\sum_{j=1}^q
P_{suc}^{(j)}$, as a function of $n$.
Comparing Fig. \ref{Fig2} with Fig. \ref{Fig3} we see that
protocol $3$ is far better than the previous two by any aspect we
might consider. First, it achieves the greatest
$\mathcal{P}_{suc}$. Indeed, for values of $n\approx 0.9$ we can
get $\mathcal{P}_{suc}\approx 0.9$, a feat unattainable by the
previous protocols. Second, it achieves its maximum
$\mathcal{P}_{suc}$ using \textit{half} the teleportations of
protocol $1$. Third, it uses much less entanglement to achieve
those highest $\mathcal{P}_{suc}$ since after the second
teleportation the entangled states employed change from
$\ket{\Phi^+_{n_j}}$ to $\ket{\Phi^+_{n_j^2}}$. This last result
is really remarkable and surprising. It means that in the
framework of MTP less entanglement at each step of the protocol is
more useful to achieve a higher $\mathcal{P}_{suc}$ than keeping
the same degree of entanglement for the quantum channel. Also,
since entanglement is a precious and difficult resource to obtain,
this property of the MTP can be really useful in practical
applications. It is worth mentioning that one interesting question
remains to be answered. Is this protocol the optimal one? For just
a few teleportations a partial analysis suggest that protocol $3$
may be the optimal one. However, no general proof, even
numerically, is available yet.

There is another property 
which is also existent in the previous two protocols. 
Looking at $\mathcal{P}_{suc}$ as a function of the number of
teleportations we see it achieves its maximal value after a small
number of steps. 
This 
is more evident the lower the entanglement of the quantum channel.
Looking at Fig. \ref{Fig3} we see that for $n<0.6$ just three
teleportations are enough to achieve the maximal
$\mathcal{P}_{suc}$. And for higher values of $n$, a few more
give the same feature. This is a 
practical property of MTP for we do not need to implement a
prohibitively large number of teleportations to get the optimal
value of $\mathcal{P}_{suc}$. One last remark. We can also look
at protocol $3$ 
as a way to correct 
errors in previous teleportations. If it is discovered that in a
previous step of the protocol an error changed the entangled state
used in the teleportation process we can correct it 
by properly choosing the right entangled state for the next
teleportation.

\section{Comparison with multiple entanglement swapping}
\label{comparison}

\begin{figure}[!ht]
\includegraphics[width=7cm]{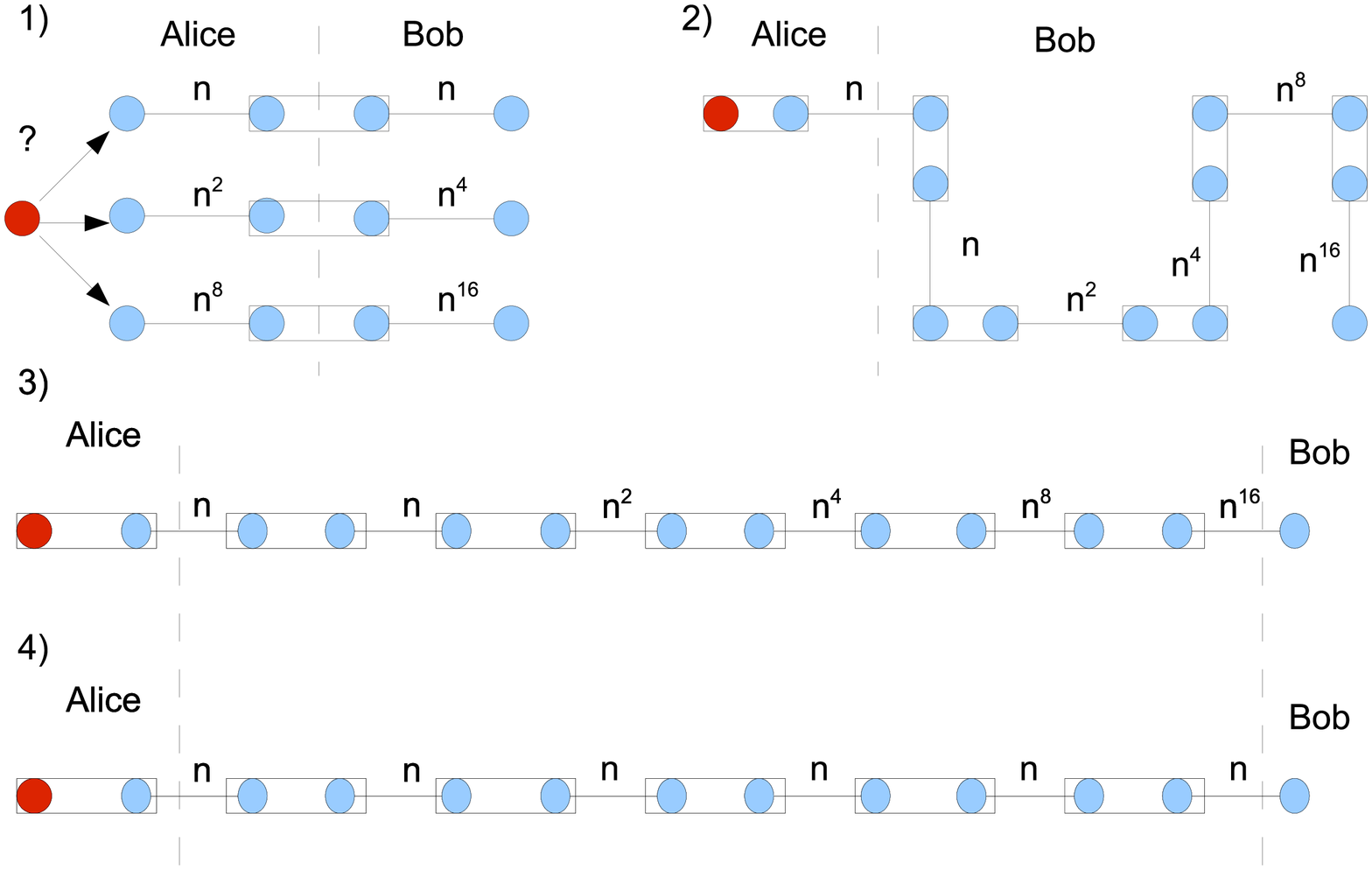}
\caption{\label{Fig6}(Color online) (1), (2), and (3) show three
possible configurations involving six PES's with decreasing
entanglement while (4) shows configuration (3) with six PES's
possessing the same entanglement. The dashed vertical lines
delimitate which qubits Alice and Bob have access to, the solid
line rectangles represent Bell measurements, and solid lines mean
entanglement between the connected qubits.}
\end{figure}

So far we have considered a ``direct approach" to teleport a qubit
using PES's. By direct we mean that we use the PES's as they are
offered to us, without any pre-processing. We have also assumed
that Bob has access to $N-1$ PES's out of a total of $N$. But we
can change this scenario in at least two ways. On the one hand we
can impose that Bob has access to only one PES. The other $N-2$
states lie between Alice and Bob. See the bottom of Fig.
\ref{Fig6}. On the other hand we can first try to extract a
maximally entangled state out of those $N$ PES's and only then
implement the usual, single-shot, teleportation protocol. See the
top-left of Fig. \ref{Fig6}, for example. Our goal in this section
is to compare the efficiencies (probabilities of success) for the
present direct protocols with the ones achieved using the multiple
entanglement swapping protocol (``swapping approach") of Ref.
\cite{Per08}, whose goal is to obtain out of $N$ PES's linking
Alice and Bob (bottom of Fig. \ref{Fig6}, for example) one
maximally entangled state (a Bell state). In this ``indirect
approach", a sequence of $N-1$ joint measurements (not only Bell
measurements) are implemented on qubits from different entangled
states (solid rectangles of Fig. \ref{Fig6}), with the hope that
at the end of the protocol the two qubits at end of the chain
become entangled.  These measurements are chosen in such a way to
maximize the probability of Alice and Bob getting a maximally
entangled two-qubit state (Bell state) at the end of the protocol.
It is this Bell state that afterwards is employed to teleport the
qubit with Alice to Bob. As will be shown, we achieve the highest
probability of success (unity fidelity teleportation) sometimes
using the direct or the swapping approach. The best strategy is
dictated by the degree of entanglement of the PES's and also by
the way they are distributed between Alice and Bob.

We start our analysis comparing the total probability of success
for protocols $2$ and $3$ against the total probability of success
for the swapping approach as giving by Eq. $D3$ of Ref.
\cite{Per08}, the best strategy for multiple swapping
teleportation. Equation $D3$ gives the probability ($P_{swap}$) of
getting one maximally entangled state out of $N$ PES's, which can
then be used to implement the usual teleportation scheme. To
derive Eq. $D3$ it is assumed that all PES's have the
\textit{same} entanglement and that Alice and Bob have access to
only one PES, as depicted at the bottom of Fig. \ref{Fig6}. In the
present notation, Eq. $D3$ reads $$P_{swap} =
1-(f_n^2-g_n^2)\sum_{j=0}^{[N/2]}f_n^{2j}g_n^{2j}\left(
\begin{array}{c}2j \\ j\end{array}\right),$$ with $[N/2]$ denoting
the integer part of $N/2$, $\left(
\begin{array}{c}2j \\ j\end{array}\right)$ meaning the binomial coefficient,
$f_n= 1/\sqrt{1 + n^2}$, and $g_n=n/\sqrt{1 + n^2}$.

\begin{figure}[!ht]
\includegraphics[width=7cm]{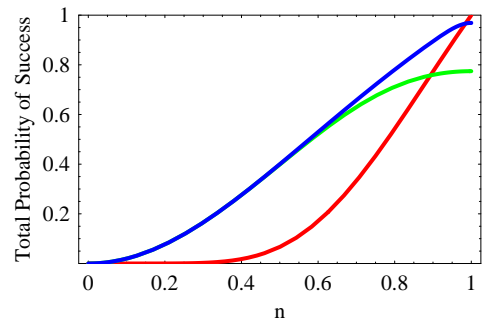}
\caption{\label{Fig4}(Color online) Upper curves represent in
ascending order $\mathcal{P}_{suc}$ for protocols $2$ (green) and
$3$ (blue) in configuration $(2)$ of Fig. \ref{Fig6}. The bottom
curve (red) gives $\mathcal{P}_{suc}$ for the swapping approach at
configuration $(4)$ of Fig. \ref{Fig6}. See text for more
details.}
\end{figure}

In our first analysis we consider for protocols $2$ and $3$ that
the $N-1$ PES's are with Bob. For protocol $2$ they all have the
same entanglement while for protocol $3$ the entanglement
decreases as explained in the previous section. (See top-right of
Fig. \ref{Fig6}.) Note that for protocol $2$ we have generalized
Bell measurements. For the swapping approach, we consider the
configuration given at the bottom of Fig. \ref{Fig6}. The results
for this scenario are illustrated in Fig. \ref{Fig4}, where we
plot the probabilities of success for $N=6$ PES's. Note that in
this situation, protocols $2$ or $3$ are superior for almost all
the range of the parameter $n$.

\begin{figure}[!ht]
\includegraphics[width=7cm]{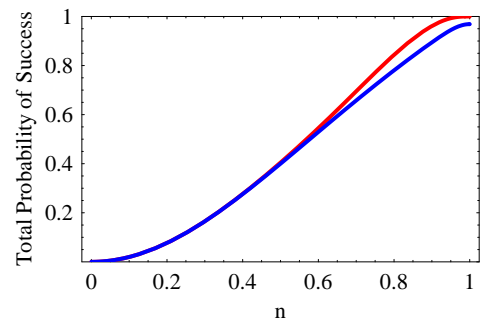}
\caption{\label{Fig5}(Color online) The upper/red curve represents
$\mathcal{P}_{suc}$ for the swapping protocol in the configuration
(1) of Fig. \ref{Fig6} and the lower/blue one $\mathcal{P}_{suc}$
for protocol $3$ at configuration (2) of Fig. \ref{Fig6}.}
\end{figure}

We now compare the swapping approach as given by configuration
$(1)$ of Fig. \ref{Fig6}, the optimal way for a swapping-based
protocol, with protocol $3$ as given by configuration $(2)$ of
Fig. \ref{Fig6}. Since we have three chances (three pairs of
PES's) for succeeding, we get for the swapping protocol
$\mathcal{P}_{suc} = S_1 + (1-S_1)S_2 + (1-S_1)(1-S_2)S_3$. Here
$S_j$, $j=1,2,3$, gives the optimal probability to obtain a
maximally entangled state out of two pairs of PES's. One can show
that \cite{Per08} $S_j=2n_j^2/(1+n_j^2)$, with $n_1=n$, $n_2=n^4$,
and $n_3=n^{16}$. Looking at Fig. \ref{Fig5} we see that in this
case the swapping protocol is slightly superior for
$n\stackrel{>}{\sim} 0.6$ while for small $n$ they both give the
same efficiencies. We should also mention that \textit{if} all the
six pairs of PES's are shared between Alice and Bob, a complete
different scenario from the ones depicted in Fig. \ref{Fig6},
entanglement concentration/filtering techniques applied
individually to all the six pairs \cite{Vid99} give a better
performance. This is true because the optimal probability to
locally concentrate a maximally entangled state from a
non-maximally pure one is $P_{con} = S_j$ \cite{Vid99}. However,
entanglement concentration can only be applied if Alice and Bob
initially do share entangled states. In the majority of the
situations studied here, though, Alice and Bob do not initially
share any entangled state and we have no choice but to rely on the
multiple teleportation or on the multiple swapping techniques.

\begin{figure}[!ht]
\includegraphics[width=6cm]{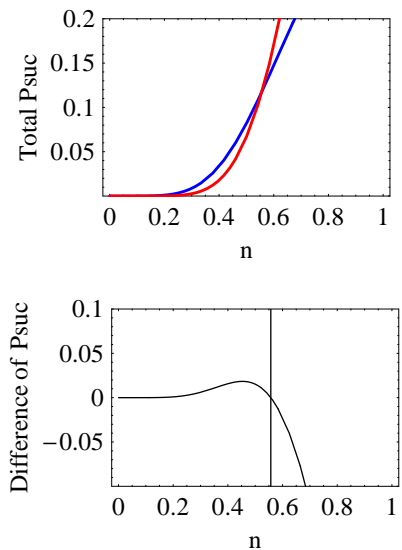}
\caption{\label{Fig7}(Color online) Top: For small values of $n$,
the upper/blue curve represents $\mathcal{P}_{suc}$ for protocol 1
while the bottom/red curve $\mathcal{P}_{suc}$ for the swapping
approach. Both cases are analyzed at configuration (4) of Fig.
\ref{Fig6}. Bottom: The vertical line (n=0.557) marks the critical
value below which the direct approach yields a better
performance.}
\end{figure}

We end this section comparing both approaches at the same
configuration, namely, configuration (4) of Fig. \ref{Fig6}. For
the direct approach we employ protocol $1$. In this scenario
$\mathcal{P}_{suc}$ for the swapping approach is given by Eq. $D3$
of Ref. \cite{Per08}, where we assume all PES's to be described by
the state $|\Phi^+_n\rangle$. For protocol $1$ $\mathcal{P}_{suc}$
is calculated considering \textit{only} those instances in which
the qubit arrives with unity fidelity at its final destination.
This always happens whenever the Bell measurements after the six
teleportations yield a balanced number of $|\Phi^{\pm}\rangle$ and
$|\Psi^{\pm}\rangle$. A simple numerical count gives $1280$
possible ways that this can happen with the total probability
being $\mathcal{P}_{suc}=20n^6/(1+n^2)^6$. Fig. \ref{Fig7} shows
$\mathcal{P}_{suc}$ for both approaches when we have six PES's. It
is interesting to note that for $n<0.557$ the direct approach is
the best choice. We have numerically checked that the lower the
number of PES's the greater the value of $n$ below which the
direct approach is the best choice. For more than $10$ PES's the
swapping protocol can be considered the best choice.
Finally, we have compared the efficiency of protocol $1$ in
configuration (4) of Fig. \ref{Fig6} against protocol $3$ in
configuration (3). We always obtained better results for protocol
$1$ in this case.

\section{Conclusion}

We have shown that the properties of the multiple teleportation
protocol (MTP) are a general feature of successive teleportations,
not being restricted to the Knill-Laflamme-Milburn (KLM) scheme.
We have also connected one formulation of MTP to the probabilistic
quantum teleportation (PQT), another approach that aims to achieve
unity fidelity teleportation via partially entangled states
(PES's). Moreover, we have presented two new MTP's that are more
efficient than the original one. Indeed, in those two new MTP's we
just need \textit{half} the number of teleportations of the
original MTP to achieve at least the same probability of success
(unity fidelity teleportation). On top of that, we have shown that
the protocol furnishing the highest probability of success
(protocol $3$) is the one requiring, surprisingly, the least
amount of entanglement for its full implementation. On the one
hand, this result may have important practical applications, since
it is known that entanglement is a difficult resource to produce
experimentally, and, on the other hand, it suggests that whenever
PES's are at stake, perhaps the best strategy to achieve a certain
goal is not the one that uses the greatest amount of entanglement.
Finally, we have compared the three MTP's here developed with the
multiple entanglement swapping approach developed in Ref.
\cite{Per08}. We have checked that either one or the other
approach furnished a better performance, depending on the amount
of entanglement available and on the way the PES's are distributed
between Alice and Bob.

\begin{acknowledgments}
The author thanks the Brazilian agency Coordena\c{c}\~ao de
Aperfei\c{c}oamento de Pessoal de N\'{\i}vel Superior (CAPES) for
funding this research.
\end{acknowledgments}

\end{document}